\documentclass[aps,preprint,showpacs,preprintnumbers,amsmath,amssymb]{revtex4}
\usepackage{amsmath,mathrsfs,amsbsy,color,graphicx,bm,amsthm,amsfonts}
\usepackage{units}
\usepackage{bbm}
\usepackage{times}
\usepackage{dcolumn}
\usepackage{mathrsfs}
\usepackage{amsmath,amssymb,epsfig}
\usepackage{amsmath}
\newcommand{\udots}{\mathinner{\mskip1mu\raise1pt\vbox{\kern7pt\hbox{.}}
\mskip2mu\raise4pt\hbox{.}\mskip2mu\raise7pt\hbox{.}\mskip1mu}}
\begin{document}
\title{Bosonic and fermionic coherence of N-partite states in the background of a dilaton black hole}
\author{ Wen-Mei Li$^1$,   Shu-Min Wu$^1$\footnote{Email: smwu@lnnu.edu.cn}}
\affiliation{$^1$ Department of Physics, Liaoning Normal University, Dalian 116029, China}


\begin{abstract}
We study the N-partite coherences of GHZ and W states for free bosonic and fermionic fields when any
$n$ observers hover near the event horizon of a Garfinkle-Horowitz-Strominger (GHS) dilaton black hole. We derive the more general analytical expressions for N-partite coherence, encompassing both physically accessible and inaccessible coherences in the context of the dilaton black hole. It has been found that the coherence of the bosonic field is greater than that of the fermionic field, while the entanglement of the fermionic field is greater than that of the bosonic field in dilaton spacetime. Additionally, the coherence of the W state is greater than that of the GHZ state, whereas the entanglement of the GHZ state is greater than that of the W state in curved spacetime. These results suggest that we should utilize suitable quantum resources and different types of particles for relativistic quantum information tasks.
\end{abstract}

\vspace*{0.5cm}
\maketitle
\section{Introduction}
Quantum coherence, one of the defining features of quantum mechanics, originates from the superposition of quantum states and underlies the fundamental phenomena of quantum interference.
Quantum coherence is a crucial quantum resource and plays a significant role in physics and quantum information processing (QIP) \cite{L111,L112,L113}. It is essential for applications  such as quantum metrology \cite{L1,L2}, quantum cryptography \cite{L3,L4,L5}, energy transport in a biological system \cite{L6}, and nanoscale thermodynamics \cite{L7,L8,L9}. Meanwhile, the framework of quantum coherence extends from single particle system to multipartite systems with several applications, such as incoherence teleportation \cite{L10} and coherence localization \cite{L11}. Quantification of quantum coherence was first proposed by Baumgratz $et$ $al.$  from the point of resource theory, introducing two typical measures: the relative entropy of coherence (REC) and the $l_{1}$ norm of coherence \cite{L12,L13,L14,L15}. Compared to the relative entropy of coherence, the $l_{1}$ norm of coherence is easier to calculate and analyze, particularly when examining the physical meaning of coherence in multipartite systems \cite{SMWL1,SMWL2}.

Black holes, formed by the gravitational collapse of sufficiently massive stars \cite{L17}, are some of the most mysterious objects in the universe. Recently, an increasing number of astronomical observations have confirmed the existence of black holes \cite{L18}. One such black hole is the Garfinkle-Horowitz-Strominger (GHS) dilaton black hole \cite{L19,L20,L21}. The dilaton black hole is predicted by superstring theory \cite{L25,L26}.  Superstring theory abandons the assumption that elementary particles are point particles, instead positing that they are one-dimensional strings. This theory is self-consistent and proposes that all particles in nature are different vibration modes of these strings. Unlike gauge theory and quantum field theory, superstring theory requires the existence of gravity, supersymmetry, and the gauge principle \cite{L25,L26,L27}. One of the notable features of superstring theory is its ability to unify gravity with other forces mediated by gauge fields. Consequently, superstring theory is considered one of the most important candidates for a theory of quantum gravity. Bipartite steering \cite{P18,P19}, entanglement \cite{P20,L24,P22}, discord \cite{L23}, quantum Fisher information \cite{P24}, and the entropic uncertainty relation \cite{P25,P26} have been studied in the context of the GHS dilaton black hole. As we know, quantum coherence is a prerequisite for steering, entanglement, and discord. However, the N-partites coherence of GHZ and W states for bosonic and fermionic fields have not been studied in the GHS dilaton black hole due to the complex calculations involved.  It is one of the motivations for studying multipartite coherence in GHS dilaton spacetime.  Previous studies have shown that fermionic entanglement is greater than bosonic entanglement in curved spacetime \cite{SMWL3,SMWL4,L30,L31,L32,L33,L34,L35,L36,L37}. However, the differences between bosonic and fermionic coherences in the GHS dilaton black hole remain unclear. It has also been demonstrated that the entanglement of GHZ state is greater than that of W state in a relativistic frame \cite{L35,L36,L37}. We aim to compare the coherences of GHZ and W states in the GHS dilaton black hole. Addressing these two questions is another motivation for our study.

In this paper, we study the N-partite coherences of GHZ and W states for bosonic and fermionic fields in the background of a GHS dilaton black hole. Initially, we assume that $N$ observers share GHZ and W states in an asymptotically flat region. Then, we consider a scenario where $N-n$ observers remain stationary in the asymptotically flat region, while $n$ observers hover near the event horizon of the GHS dilaton black hole. We aim to derive more general analytical expressions for N-partite coherences  that include all physically accessible and inaccessible coherences in GHS dilaton black hole. We find that the coherence of the bosonic field is larger than that of the fermionic field, while the entanglement of the fermionic field is larger than that of the bosonic field in relativistic frame \cite{L30,L31,L32,L33,L34,L35,L36,L37}; we also find that the coherence of W state is larger than that of GHZ state, while the entanglement of GHZ state is larger than that of W state in relativistic frame \cite{L36}.
Therefore, we should choose suitable quantum resources and suitable type of particles to deal with the relativistic quantum information tasks.

The structure of the paper is as follows. In Sec. II, we introduce the quantization of bosonic and fermionic fields in GHS dilaton spacetime.  In Sec. III, we discuss N-partite coherence of bosonic and fermionic fields in the GHS dilaton black hole.  In Sec. IV, we compare coherence with entanglement of bosonic and fermionic fields in GHS dilaton spacetime. The last section is devoted to the summary.

\section{The quantization of field in GHS dilaton spacetime \label{GSCDGE}}
The metric of a GHS black hole spacetime becomes \cite{L19,L20}
\begin{eqnarray}\label{S1}
ds^{2}=-\bigg(\frac{r-2M}{r-2\widetilde{\mathcal{D}}M}\bigg)dt^{2}+\bigg(\frac{r-2M}{r-2\widetilde{\mathcal{D}}M}\bigg)^{-1}dr^{2}+r(r-2\widetilde{\mathcal{D}}M)d\Omega^{2},
\end{eqnarray}
where $\widetilde{\mathcal{D}}$ and $M$ are parameters related to the dilaton field and the mass of the black hole, respectively. The relationship among $\widetilde{\mathcal{D}}$, the charge $Q$, and $M$ is described by the equation $\widetilde{\mathcal{D}}=Q^{2}/2M^{2}$. In this paper, we consider $G=c=\hbar=\kappa_{B}=1$.
\subsection{Bosonic field}

The couple massive scalar filed of general perturbation equation in this dilaton spacetime is given as \cite{L26}
\begin{eqnarray}\label{S2}
\frac{1}{\sqrt{-g}}\partial_{\mu}(\sqrt{-g}g^{\mu\nu}\partial_{\nu})\Psi-(\mu+\xi R)\Psi=0,
\end{eqnarray}
where $\mu$ represents the mass of the particle, $R$ is the Ricci scalar curvature, and $\psi$ is the scalar filed. The coupling between the gravitation filed and the scalar filed is represented by the term $\xi R \psi$, where $\xi$ is a numerical coupling factor. The normal mode solution can be expressed as
\begin{eqnarray}\label{S3}
\Psi_{\omega lm}=\frac{1}{h(r)}\chi_{\omega l}(r)Y_{lm}(\theta,\varphi)e^{-i\omega t},
\end{eqnarray}
where $Y_{lm}(\theta,\varphi)$ represents a scalar spherical harmonic on the unit two-sphere and $h(r)=\sqrt{r(r-2D)}$ with $D=Q^{2}/2M$. We can easily obtain the radial equation
\begin{eqnarray}\label{S4}
\frac{d^{2}\chi_{\omega l}}{dr^{2}_{\ast}}+[\omega^{2}-V(r)]\chi_{\omega l}=0,
\end{eqnarray}
and
\begin{eqnarray}\label{S5}
V(r)=\frac{f(r)}{h(r)}\frac{d}{dr}\bigg[f(r)\frac{dh(r)}{dr}\bigg]+\frac{f(r)l(l+1)}{h^{2}(r)}+f(r)\bigg[\mu^{2}+\frac{2\xi D^{2}(r-2M)}{r^{2}(r-2D)^{3}}\bigg],
\end{eqnarray}
where $f(r)=(r-2M)/(r-2D)$ and the tortoise coordinate $r_{\ast}$ defined  as $dr_{\ast}=dr/f(r)$ .

Solving Eq.(\ref{S4}) near the event horizon, one can obtain the incoming wave function, which is analytic everywhere in the spacetime manifold
\begin{eqnarray}\label{S6}
\Psi_{in,\omega lm}=e^{-i\omega v}Y_{lm}(\theta,\varphi),
\end{eqnarray}
and the outgoing wave functions for the outside and inside region of the event horizon are given by
\begin{eqnarray}\label{S7}
\Psi_{out,\omega lm}(r>r_{+})=e^{-i\omega u}Y_{lm}(\theta,\varphi),
\end{eqnarray}
\begin{eqnarray}\label{S8}
\Psi_{out,\omega lm}(r<r_{+})=e^{i\omega u}Y_{lm}(\theta,\varphi),
\end{eqnarray}
where $u=t-r_{\ast}$ and $v=t+r_{\ast}$. Eqs.(\ref{S7}) and (\ref{S8}) are separately analytic outside and inside the event horizon. Therefore, they become a completely orthogonal family.

The generalized light-like Kruskal coordinates are defined as \cite{L24,P18}
\begin{eqnarray}\label{S9}
u&=&-4(M-D)\ln[-U/(4M-4D)],\notag\\v&=&4(M-D)\ln[V/(4M-4D)],\mathrm{if}\;r>r_{+},\notag\\
u&=&-4(M-D)\ln[U/(4M-4D)],\notag\\v&=&4(M-D)\ln[V/(4M-4D)],\mathrm{if}\;r<r_{+}.
\end{eqnarray}
Eqs.(\ref{S7}) and (\ref{S8}) can be rewritten as
\begin{eqnarray}\label{S10}
\Phi_{out,\omega lm}(r>r_{+})=e^{4(M-D)i\omega\ln[U/(4M-4D)]}Y_{lm}(\theta,\varphi),
\end{eqnarray}
\begin{eqnarray}\label{S11}
\Phi_{out,\omega lm}(r<r_{+})=e^{-4(M-D)i\omega\ln[-U/(4M-4D)]}Y_{lm}(\theta,\varphi).
\end{eqnarray}
Employing the formula $-1=e^{i\pi}$ and making Eq.(\ref{S10}) analytic in the lower half-plane of $U$, a complete basis for positive energy $U$ modes can be expressed as
\begin{eqnarray}\label{S12}
\Phi_{I,\omega lm}=e^{2\pi\omega(M-D)}\Phi_{out,\omega lm}(r>r_{+})
+e^{-2\pi\omega(M-D)}\Phi_{out,\omega lm}^{\ast}(r<r_{+}),
\end{eqnarray}
\begin{eqnarray}\label{S13}
\Phi_{II,\omega lm}=e^{-2\pi\omega(M-D)}\Phi_{out,\omega lm}^{\ast}(r>r_{+})
+e^{2\pi\omega(M-D)}\Phi_{out,\omega lm}(r<r_{+}).
\end{eqnarray}
Eqs.(\ref{S12}) and (\ref{S13}) are analytic for all real $U$ and $V$, which form a complete basis for positive frequency modes.  Therefore, we can also use $\Phi_{I,\omega lm}$ and $\Phi_{II,\omega lm}$ to quantize the quantum filed in Kruskal spacetime.

By applying second quantization to the field in the exterior of the dilaton black hole \cite{P18,L134,L135,L136,L40}, we can derive the Bogoliubov transformation for the creation and annihilation operators in both dilaton and Kruskal spacetime
\begin{eqnarray}\label{S14}
a_{K,\omega lm}^{B,\dag}=\frac{1}{\sqrt{1-e^{-8\pi\omega(M-D)}}}b_{out,\omega lm}^{B,\dag}
-\frac{1}{\sqrt{e^{8\pi\omega(M-D)}-1}}b_{in,\omega lm}^{B},
\end{eqnarray}
\begin{eqnarray}\label{S15}
a_{K,\omega lm}^{B}=\frac{1}{\sqrt{1-e^{-8\pi\omega(M-D)}}}b_{out,\omega lm}^{B}-\frac{1}{\sqrt{e^{8\pi\omega(M-D)}-1}}b_{in,\omega lm}^{B,\dag},
\end{eqnarray}
where $a_{K,\omega lm}^{B,\dagger}$ and $a_{K,\omega lm}^{B}$ are the creation operator and annihilation operator, respectively, acting on the Kruskal vacuum of the exterior region, $b_{in,\omega lm}^{B,\dagger}$ and $b_{in,\omega lm}^{B}$ are the creation operator and annihilation operator, respectively, acting on the dilaton vacuum of the interior region of the dilaton black hole, and $b_{out,\omega lm}^{B,\dagger}$ and $b_{out,\omega lm}^{B}$ are the creation operator and annihilation operator, respectively, acting on the dilaton vacuum of the exterior region. Therefore, the Kruskal vacuum $|0\rangle_{K}^{B}$ can be defined as
\begin{eqnarray}\label{S16}
a_{K,\omega lm}^{B}|0\rangle_{K}^{B}=0.
\end{eqnarray}
After appropriately normalizing the state vector, the Kruskal vacuum of bosonic field in dilaton spacetime is a maximally entangled two-mode squeezed state \cite{L137,L138}
\begin{eqnarray}\label{S17}
|0\rangle_{K}^{B}=\sqrt{1-e^{-8\pi\omega(M-D)}}\sum^{\infty}_{n=0} e^{-4n\pi\omega(M-D)}|n\rangle_{out}^{B}|n\rangle_{in}^{B},
\end{eqnarray}
and the first excited state of boconic field reads
\begin{eqnarray}\label{S18}
|1\rangle_{K}^{B}=a_{K,\omega lm}^{\dagger}|0\rangle _{K}^{B}=
\big[1-e^{-8\pi\omega(M-D)}\big]\sum^{\infty}_{n=0}\sqrt{n+1} e^{-4n\pi\omega(M-D)}|n+1\rangle_{out}^{B}|n\rangle_{in}^{B},
\end{eqnarray}
where $B$ represents the bosonic field, $\{|n\rangle_{out}\}$ and $\{|n\rangle_{in}\}$ are the orthonormal bases for the outside and inside regions of the event horizon, respectively. For an observer outside the dilaton black hole, it is necessary to trace over the modes in the interior region, as they have no access to the information within this causally disconnected region. Consequently, we can derive the Hawking radiation spectrum
\begin{eqnarray}\label{S19}
N^{B}_{\omega}=\frac{1}{e^{8\pi\omega(M-D)}-1}.
\end{eqnarray}
Eq.(\ref{S19}) shows that an observer in the exterior of the GHS dilaton black hole detects a thermal Bose-Einstein distribution of particles when traversing the Kruskal vacuum.

\subsection{Fermionic field}

Similar to the bosonic field, one obtain the Bogoliubov relations between the Kruskal and dilaton operators of ferminic field as \cite{P19,P32}
\begin{eqnarray}\label{S20}
a^{F,\dag}_{K,\omega lm}=\frac{1}{\sqrt{e^{-8\pi\omega(M-D)}+1}}a^{F,\dag}_{out,\omega lm}-\frac{1}{\sqrt{e^{8\pi\omega(M-D)}+1}}b_{in,\omega lm}^{F},
\end{eqnarray}
\begin{eqnarray}\label{S21}
a^{F}_{K,\omega lm}=\frac{1}{\sqrt{e^{-8\pi\omega(M-D)}+1}}a_{out,\omega lm}^{F}-\frac{}{\sqrt{e^{8\pi\omega(M-D)}+1}}b^{F,\dag}_{in,\omega lm}.
\end{eqnarray}
Therefore, the  Kruskal vacuum and excited states of the fermionic field in dilaton spacetime can
be expressed as
\begin{eqnarray}\label{S22}
|0\rangle_{K}^{F}=\frac{1}{\sqrt{e^{-8\pi\omega(M-D)}+1}}|0_{K}\rangle_{out}^{F}|0_{-K}\rangle_{in}^{F}+\frac{1}{\sqrt{e^{8\pi\omega(M-D)}+1}}|1_{K}\rangle_{out}^{F}|1_{-K}\rangle_{in}^{F},
\end{eqnarray}
and
\begin{eqnarray}\label{S23}
|1\rangle_{K}^{F}=|1_{K}\rangle_{out}^{F}|0_{-K}\rangle_{in}^{F},
\end{eqnarray}
where $F$ represents the fermionic field. Through similar solution of the bosonic field, one can obtain the Hawking radiation spectrum  of the fermionic field  as
\begin{eqnarray}\label{S24}
N^{F}_{\omega}=\frac{1}{e^{8\pi\omega(M-D)}+1}.
\end{eqnarray}
This equation indicates that the observer outside the event horizon
detects a thermal Fermi-Dirac distribution of particles. Through Eqs.(\ref{S19}) and (\ref{S24}), we can see that these two types of distributions will lead to gravitational effects in dilaton spacetime that have different effects on the coherence and entanglement of the bosonic and fermionic fields.


\section{N-partite coherence in GHS dilaton black hole  \label{GSCDGE}}
We assume that $N$ $(N\geq3)$ observers share GHZ state
\begin{eqnarray}\label{S25}
GHZ^{B/F}_{123\ldots N}=\frac{1}{\sqrt{2}}(|00\ldots00\rangle+|11\ldots11\rangle),
\end{eqnarray}
and W state
\begin{eqnarray}\label{S26}
W^{B/F}_{123\ldots N}&=&\frac{1}{\sqrt{N}}(|10\ldots 00\rangle+|01\ldots 00\rangle+\cdots+|00\ldots 01\rangle),
\end{eqnarray}
at the same point in the asymptotically flat region of the dilaton black hole. Here, we use $B$ and $F$ to denote boson and fermion, respectively. After sharing their own qubit, $n$ $(1\leq n\leq N)$ observers hover near the event horizon of the dilaton black hole, while the remaining $N-n$ observers stay stationary in an asymptotically flat region.
\subsection{N-partite coherences of GHZ and W states for bosonic field}
We use Eqs.(\ref{S17}) and (\ref{S18}) to rewrite Eq.(\ref{S25}) in the background of the dilaton black hole
\begin{eqnarray}\label{S27}
GHZ^{B}_{123\ldots N+n}&=&\frac{1}{\sqrt{2}}\bigg\{(1-e^{-8\pi\omega(M-D)})^{\frac{n}{2}}\overbrace{(|0\rangle_{N}|0\rangle_{N-1}\cdots|0\rangle_{n+1})}^{|\bar{0}\rangle}\notag\\
&&\bigotimes^{n}_{i=1}\bigg[\sum^{\infty}_{m_{i}=0}e^{-4\pi m_{i}\omega(M-D)}|m_{i}\rangle_{out}|m_{i}\rangle_{in}\bigg]\notag\\
&&+(1-e^{-8\pi\omega(M-D)})^{n}\overbrace{(|1\rangle_{N}|1\rangle_{N-1}\cdots|1\rangle_{n+1})}^{|\bar{1}\rangle}\notag\\
&&\bigotimes^{n}_{i=1}\bigg[\sum^{\infty}_{m_{i}=0}e^{-4\pi m_{i}\omega(M-D)}\sqrt{m+1_{i}}|m+1_{i}\rangle_{out}|m_{i}\rangle_{in}\bigg]\bigg\},
\end{eqnarray}
where $|\bar{0}\rangle=|0\rangle_{n+1}|0\rangle_{n+2}\cdots|0\rangle_{N}$ and $|\bar{1}\rangle=|1\rangle_{n+1}|1\rangle_{n+2}\cdots|1\rangle_{N}$. In order to investigate physically accessible and inaccessible coherences, we consider a general system $\rho^{B,GHZ}_{N-n,p,q} (p+q=n)$, which consists of $N-n$ modes in the asymptotically flat region, physically accessible $p$ modes outside the event horizon, and physically inaccessible $q$ modes inside the event horizon. The density operator $\rho^{B,GHZ}_{N-n,p,q}$ can be expressed as
\begin{eqnarray}\label{S28}
\rho^{B}(GHZ)&=&\frac{1}{2}\bigg\{\big[(1-e^{-8\pi\omega(M-D)})\big]^{n}\prod_{i=1}^{n}\big[\sum^{\infty}_{m_{i}=0}\alpha_{m_i}^{2}\big]|\overline{0}\rangle\langle\overline{0}|\big[\bigotimes^{p}_{i=1}|m_{i}\rangle _{out}\langle m_{i}|\big]\big[\bigotimes^{n}_{i=p+1}|m_{i}\rangle _{in}\langle m_{i}|\big]\notag\\
&+&\big[(1-e^{-8\pi\omega(M-D)})\big]^{\frac{3n}{2}}\prod_{i=1}^{p}\big[\sum^{\infty}_{m_{i}=0}\alpha_{m_i}\beta_{m_i}\big]\prod_{j=1}^{q}\big[\sum^{\infty}_{m_{j}=0}\beta_{m_j}\gamma_{m_j}\big]|\overline{0}\rangle\langle\overline{1}|\notag\\
&\times&\big[\bigotimes^{p}_{i=1}|m_{i}\rangle _{out}\langle m+1_{i}|\big]\big[\bigotimes^{q}_{j=1}|m+1_{j}\rangle _{in}\langle m_{j}|\big]+\big[(1-e^{-8\pi\omega(M-D)})\big]^{\frac{3n}{2}}\notag\\
&\times&\prod_{i=1}^{p}\big[\sum^{\infty}_{m_{i}=0}\alpha_{m_i}\beta_{m_i}\big]\prod_{j=1}^{q}\big[\sum^{\infty}_{m_{j}=0}\beta_{m_j}\gamma_{m_j}\big]\overline{1}\rangle\langle\overline{0}|\big[\bigotimes^{p}_{i=1}|m+1_{i}\rangle _{out}\langle m_{i}|\big]\notag\\
&\times&\big[\bigotimes^{q}_{j=1}|m_{j}\rangle _{in}\langle m+1_{j}|\big]+\big[(1-e^{-8\pi\omega(M-D)})\big]^{2n}\prod_{i=1}^{n}\big[\sum^{\infty}_{m_{i}=0}\beta_{m_i}^{2}\big]|\overline{1}\rangle\langle\overline{1}|\notag\\
&\times&\big[\bigotimes^{p}_{i=1}|m+1_{i}\rangle _{out}\langle m+1_{i}|\big]\big[\bigotimes^{n}_{i=p+1}|m_{i}\rangle _{in}\langle m_{i}|\big]\bigg\},
\end{eqnarray}
where
\begin{equation}\nonumber
\begin{aligned}
\alpha_{m_i}=e^{-4m_{i}\pi\omega(M-D)}, \notag\\
\beta_{m_i}=\sqrt{m+1_{i}}e^{-4m_{i}\pi\omega(M-D)}, \notag\\
\gamma_{m_i}=e^{-4(m+1_{i})_{i}\pi\omega(M-D)}.
\end{aligned}
\end{equation}

In this paper, we use the $l_{1}$ norm to measure the coherence in curved spacetime. The $l_{1}$ norm of coherence  is defined as the sum of the absolute values of all the off-diagonal elements in the system's density matrix
\begin{eqnarray}\label{S29}
C(\rho)=\sum_{i\neq j}|\rho_{i,j}|.
\end{eqnarray}
Employing the Eq.(\ref{S29}), we obtain the analytical expression of N-partite coherence of GHZ state for bosonic field in curved spacetime as
\begin{eqnarray}\label{S30}
C^{B}(GHZ)=&\bigg[&(1-e^{-8\pi\omega(M-D)})^{\frac{3}{2}}\sum^{\infty}_{i=0}\sqrt{i+1}(e^{-4m\pi\omega(M-D)})^{2i}\bigg]^{p}\notag\\
&\bigg[&(1-e^{-8\pi\omega(M-D)})^{\frac{3}{2}}\sum^{\infty}_{i=0}\sqrt{i+1}(e^{-4m\pi\omega(M-D)})^{2i+1}\bigg]^{q}.
\end{eqnarray}

Similar to GHZ state of bosonic field, N-partite W state  in terms of dilaton modes, can be rewritten as
\begin{eqnarray}\label{S31}
W^{B}_{123\ldots N+n}
&=&\frac{1}{\sqrt{N}}\bigg\{(1-e^{-8\pi\omega(M-D)})^{\frac{n}{2}}|1\rangle_{_{N}}|0\rangle_{_{N-1}}\cdots|0\rangle_{_{n+1}}\big[\sum^{\infty}_{m_{i}=0}\bigotimes^{n}_{i=1}\notag\\
&\times&\alpha_{i}|m_{i}\rangle_{out}|m_{i}\rangle_{in}\big]\notag\\
&+&(1-e^{-8\pi\omega(M-D)})^{\frac{n}{2}}|0\rangle_{_{N}}|1\rangle_{_{N-1}}\cdots|0\rangle_{_{n+1}}\big[\sum^{\infty}_{m_{i}=0}\bigotimes^{n}_{i=1}\notag\\
&\times&\alpha_{i}|m_{i}\rangle_{out}|m_{i}\rangle_{in}\big]+\cdots+(1-e^{-8\pi\omega(M-D)})^{\frac{n+1}{2}}|0\rangle_{_{N}}|0\rangle_{_{N-1}}\cdots\notag\\
&\times&|0\rangle_{_{n+1}}\big[\sum^{\infty}_{m_{i}=0}\bigotimes^{n}_{i=2}\alpha_{i}|m_{i}\rangle_{out}|m_{i}\rangle_{in}\big]\notag\\
&\times&\big[\sum^{\infty}_{m_{1}=0}\beta_{m_1}|m+1_{1}\rangle_{out}|m_{1}\rangle_{in}\big]\bigg\}.
\end{eqnarray}
Since the density operator of the W state for  bosonic field is complex, we do not attempt to write it out explicitly for simplicity. Fortunately, we can obtain the analytical expression of the coherence $C^{B}(W)$ between the physically accessible and inaccessible modes as
\begin{eqnarray}\label{S32}
C^{B}(W)&=&\frac{1}{N}\bigg\{p(p-1)\bigg[(1-e^{-8\pi\omega(M-D)})^{\frac{3}{2}}\sum^{\infty}_{i=0}(e^{-4m\pi\omega(M-D)})^{2i}\sqrt{i+1}\bigg]^{2}\notag\\
&+&q(q-1)\bigg[(1-e^{-8\pi\omega(M-D)})^{\frac{3}{2}}\sum^{\infty}_{i=0}(e^{-4m\pi\omega(M-D)})^{2i+1}\sqrt{i+1}\bigg]^{2}\notag\\
&+&2p(N-p-q)\bigg[(1-e^{-8\pi\omega(M-D)})^{\frac{3}{2}}\sum^{\infty}_{i=0}(e^{-4m\pi\omega(M-D)})^{2i}\sqrt{i+1}\bigg]\notag\\
&+&2q(N-p-q)\bigg[(1-e^{-8\pi\omega(M-D)})^{\frac{3}{2}}\sum^{\infty}_{i=0}(e^{-4m\pi\omega(M-D)})^{2i+1}\sqrt{i+1}\bigg]\notag\\
&+&(N-p-q)(N-p-q-1)\bigg\}.
\end{eqnarray}
From Eqs.(\ref{S30}) and (\ref{S32}),  we can see that for $q=0$, $C^{B}(GHZ)$ and $C^{B}(W)$ denote physically accessible coherences of GHZ and W states, respectively. Conversely,  when $p=0$, $C^{B}(GHZ)$ and $C^{B}(W)$ denote the completely inaccessible coherences of the GHZ and W states, respectively. We also see that N-partite coherence of GHZ state for bosonic field is independent of the number of initial particle $N$, while N-partite coherence of W state for bosonic field depends on the number of initial particle $N$.

\begin{figure}
\begin{minipage}[t]{0.5\linewidth}
\centering
\includegraphics[width=3.0in,height=5.2cm]{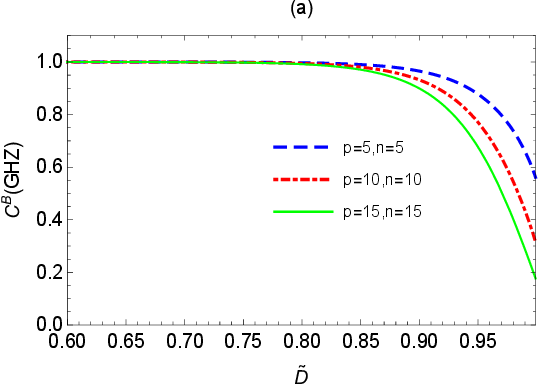}
\label{fig1a}
\end{minipage}%
\begin{minipage}[t]{0.5\linewidth}
\centering
\includegraphics[width=3.0in,height=5.2cm]{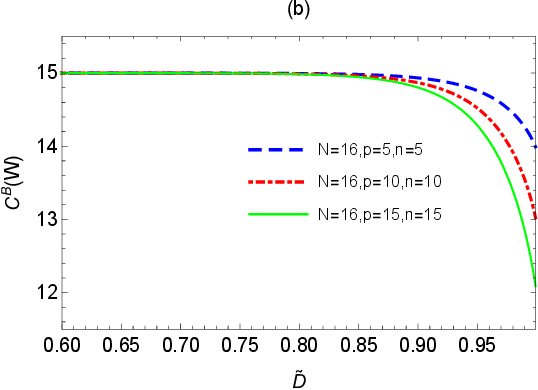}
\label{fig1b}
\end{minipage}%

\begin{minipage}[t]{0.5\linewidth}
\centering
\includegraphics[width=3.0in,height=5.2cm]{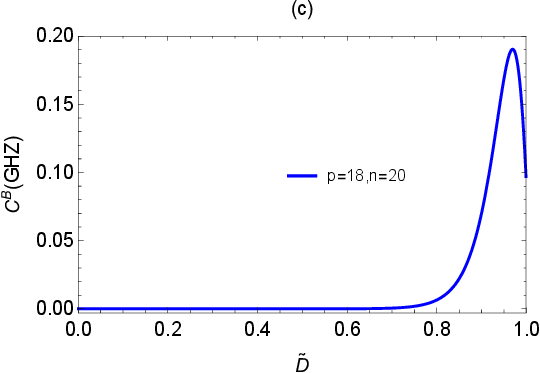}
\label{fig1c}
\end{minipage}%
\begin{minipage}[t]{0.5\linewidth}
\centering
\includegraphics[width=3.0in,height=5.2cm]{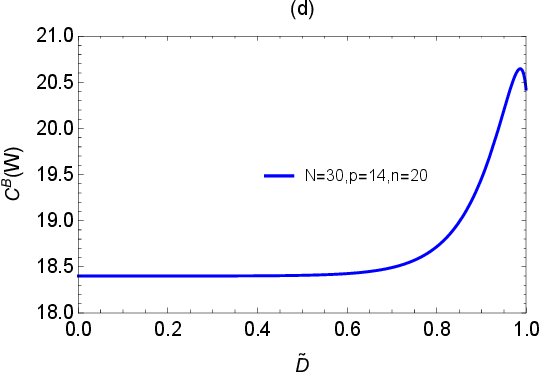}
\label{fig1d}
\end{minipage}%

\begin{minipage}[t]{0.5\linewidth}
\centering
\includegraphics[width=3.0in,height=5.2cm]{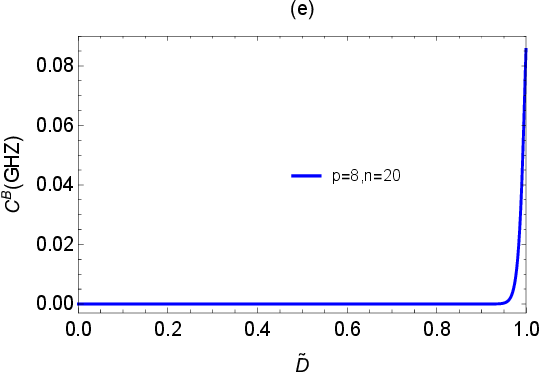}
\label{fig1e}
\end{minipage}%
\begin{minipage}[t]{0.5\linewidth}
\centering
\includegraphics[width=3.0in,height=5.2cm]{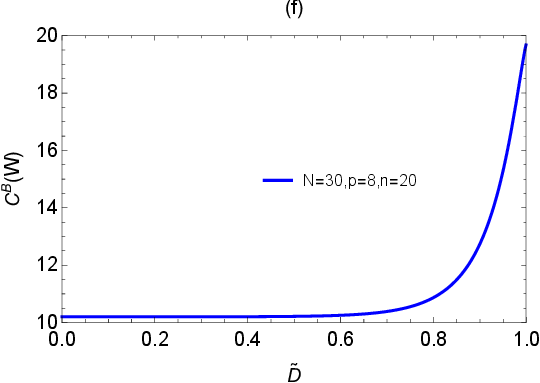}
\label{fig1f}
\end{minipage}%
\caption{Quantum coherences $C^{B}(GHZ)$ and $C^{B}(W)$ of bosonic field as a function of the dilaton $\widetilde{\mathcal{D}}$ for different $N$, $p$, $n$, where we have fixed $M=\omega=1$.}
\label{Fig1}
\end{figure}

Fig.\ref{Fig1} shows how the dilaton $\widetilde{\mathcal{D}}$ of the black hole influences the coherences of GHZ and W states for bosonic field. From Fig.\ref{Fig1}(a) and (b), we can observe the physically accessible coherence $(p=n)$ of bosonic field monotonically decreases to a fixed value with the increase of the dilaton $\widetilde{\mathcal{D}}$, which means that N-partite coherence can survive in the dilaton black hole, while bosonic steering experiences catastrophic behavior, such as "sudden death" \cite{P18}. For the fixed $N$, the physically accessible coherence decreases monotonically with the increase of the number of bosons $n$ near the event horizon of the dilaton black hole. In Fig.\ref{Fig1}(c) and (d), we also observe that the inaccessible coherence $(p=18,n=20)$ initially increases to a maximum before decreasing to a fixed value as the dilaton $\widetilde{\mathcal{D}}$ increases. However, the inaccessible coherence $(p=8,n=20)$ monotonically increases with increasing dilaton $\widetilde{\mathcal{D}}$ in Fig.\ref{Fig1}(e) and (f). In other words, the physically inaccessible N-partite coherence increases either monotonically or non-monotonically with the increase of the dilaton $\widetilde{\mathcal{D}}$, depending on the relative numbers between the accessible and the inaccessible modes. In addition, the coherence of W state is always bigger than that of GHZ state.
\begin{figure}
\begin{minipage}[t]{0.5\linewidth}
\centering
\includegraphics[width=3.0in,height=5.2cm]{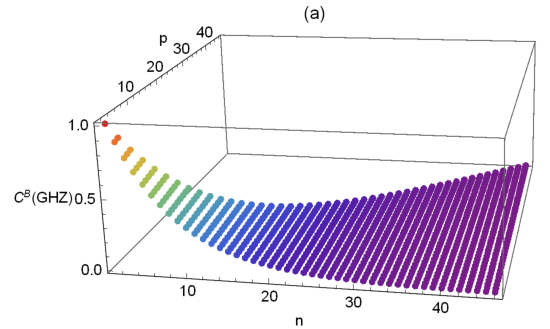}
\label{fig2a}
\end{minipage}%
\begin{minipage}[t]{0.5\linewidth}
\centering
\includegraphics[width=3.0in,height=5.2cm]{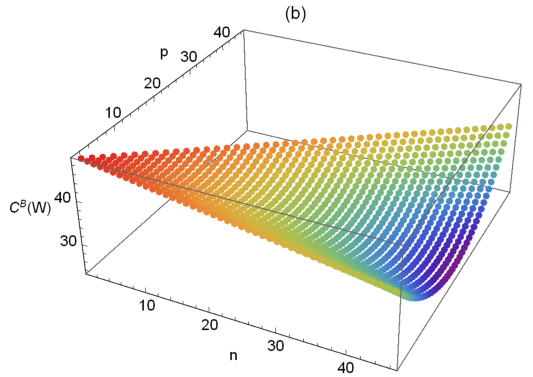}
\label{fig2b}
\end{minipage}%
\caption{Quantum coherences of GHZ and W states for bosonic field as functions of $p$ and $n$ for fixed $M=\omega=1$ and $\widetilde{\mathcal{D}}\longrightarrow1$.}
\label{Fig2}
\end{figure}

In Fig.\ref{Fig2}, we study how coherences of GHZ and W states for bosonic field depends on $p$ and $n$ in the limit of $D\rightarrow M$, corresponding to the case of an extreme black hole. From Fig.\ref{Fig2}, we can see that the diagonal lines ($p=n$) correspond to the physically accessible coherence that monotonically decreases with the growth of $p$. We can also see that the coherence of GHZ state for any $p$ first decreases and then approaches zero with the increase of $n$. However, the coherence of W state monotonically decreases to a fixed value with the increase of $n$. For fixed $n$, with increasing $p$, the coherence of GHZ state remains unchanged, while the coherence of W state initially decreases monotonically to a minimum and then increases to a fixed value. When $D\rightarrow M$, we can obtain
\begin{equation}\nonumber
\begin{aligned}
&\lim_{ D\rightarrow M}\bigg[(1-e^{-8\pi\omega(M-D)})^{\frac{3}{2}}\sum^{\infty}_{i=0}\sqrt{i+1}(e^{-4m\pi\omega(M-D)})^{2i}\bigg]&\notag\\
&=\lim_{ D\rightarrow M}\bigg[(1-e^{-8\pi\omega(M-D)})^{\frac{3}{2}}\sum^{\infty}_{i=0}\sqrt{i+1}(e^{-4m\pi\omega(M-D)})^{2i+1}\bigg]&=\frac{\sqrt{\pi}}{2}.
\end{aligned}
\end{equation}
Therefore, from Eq.(\ref{S30}), the coherence of GHZ state for bosonic field is
\begin{equation}\nonumber
\begin{aligned}
C^{B}(GHZ)=\lim_{ D\rightarrow M}\bigg[(1-e^{-8\pi\omega(M-D)})^{\frac{3}{2}}\sum^{\infty}_{i=0}\sqrt{i+1}(e^{-4m\pi\omega(M-D)})^{2i}\bigg]^{n}=\big(\frac{\sqrt{\pi}}{2}\big)^{n}.
\end{aligned}
\end{equation}

\subsection{N-partite coherences of GHZ and W states for fermionic field}
Similar to GHZ state of bosonic field, we use dilaton modes given by Eqs.(\ref{S22}) and (\ref{S23}) to rewrite Eq.(\ref{S25}) of fermionic field
\begin{eqnarray}\label{S33}
GHZ^{F}_{123\ldots N}&&=\frac{1}{\sqrt{2}}\bigg\{\bigotimes^{n}_{i=1}\overbrace{\big(|0\rangle_{N}|0\rangle_{N-1}\cdots|0\rangle_{n+1}\big)}^{|\bar{0}\rangle}\big[(e^{-8\pi\omega(M-D)}+1)^{-\frac{1}{2}}|0_{i}\rangle_{out}|0_{i}\rangle_{in}\notag\\
&&+(e^{8\pi\omega(M-D)}+1)^{-\frac{1}{2}}|1_{i}\rangle_{out}|1_{i}\rangle_{in}\big]+\bigotimes^{n}_{i=1}\overbrace{\big(|1\rangle_{N}|1\rangle_{N-1}\cdots|1\rangle_{n+1}\big)}^{|\bar{1}\rangle}\notag\\
&&\big[|1_{i}\rangle_{out}|0_{i}\rangle_{in}\big]\bigg\}.
\end{eqnarray}
We also consider a general system $\rho^{F}_{N-n,p,q}$ $(p+q=n)$ of fermionic field, which can be expressed as
\begin{eqnarray}\label{S34}
\rho^{F}(GHZ)&=&\frac{1}{2}(\frac{1}{\sqrt{e^{-8\pi\omega(M-D)}+1}})^{2n}|\overline{0}\rangle\langle\overline{0}|[\bigotimes^{p}_{i=1}(|0_{i}\rangle_{out}\langle0_{i}|)]\times[\bigotimes^{q}_{j=1}(|0_{j}\rangle_{in}\langle0_{j}|)]\notag\\
&+&\frac{1}{2}(\frac{1}{\sqrt{e^{-8\pi\omega(M-D)}+1}})^{2n-2}(\frac{1}{\sqrt{e^{8\pi\omega(M-D)}+1}})^{2}|\overline{0}\rangle\langle\overline{0}|\bigg\{\sum^{p}_{m=1}[(|1_{m}\rangle_{out}\langle1_{m}|)\notag\\
&\times&\bigotimes^{p}_{i=1,i\neq m}(|0_{i}\rangle_{out}\langle0_{i}|)\bigotimes^{q}_{j=1}(|0_{j}\rangle_{in}\langle0_{j}|)]+\sum^{q}_{m=1}[(|1_{m}\rangle_{in}\langle1_{m}|)\bigotimes^{q}_{j=1,j\neq m}(|0_{j}\rangle_{in}\langle0_{j}|)\notag\\
&\times&\bigotimes^{p}_{i=1}(|0_{i}\rangle_{out}\langle0_{i}|)]\bigg\}+\frac{1}{2}(\frac{1}{\sqrt{e^{-8\pi\omega(M-D)}+1}})^{2n-4}(\frac{1}{\sqrt{e^{8\pi\omega(M-D)}+1}})^{4}|\overline{0}\rangle\langle\overline{0}|\notag\\
&\times&\bigg\{\sum^{p}_{m=2}\sum^{m-1}_{z=1}[(|1_{m}\rangle_{out}\langle1_{m}|)(|1_{z}\rangle_{out}\langle1_{z}|)\bigotimes^{p}_{i=1,i\neq z,m}(|0_{i}\rangle_{out}\langle0_{i}|)\bigotimes^{q}_{j=1}(|0_{j}\rangle_{in}\langle0_{j}|)]\notag\\
&+&\sum^{q}_{m=2}\sum^{m-1}_{z=1}[(|1_{z}\rangle_{in}\langle1_{z}|)(|1_{m}\rangle_{in}\langle1_{m}|)\bigotimes^{q}_{j=1,j\neq z,m}(|0_{j}\rangle_{in}\langle0_{j}|)\bigotimes^{p}_{i=1}(|0_{i}\rangle_{out}\langle0_{i}|)]\notag\\
&+&\sum^{p}_{z=1}\sum^{q}_{m=1}[(|1_{z}\rangle_{out}\langle1_{z}|)\bigotimes^{p}_{i=1,i\neq z}(|0_{i}\rangle_{out}\langle0_{i}|)(|1_{m}\rangle_{in}\langle1_{m}|)\bigotimes^{q}_{j=1,j\neq m}(|0_{j}\rangle_{in}\langle0_{j}|)]\bigg\}\notag\\
&+&\cdots+\frac{1}{2}(\frac{1}{\sqrt{e^{8\pi\omega(M-D)}+1}})^{2n}|\overline{0}\rangle\langle\overline{0}|\bigotimes^{p}_{i=1}(|1_{i}\rangle_{out}\langle1_{i}|)\bigotimes^{q}_{j=1}(|1_{j}\rangle_{in}\langle1_{j}|)\notag\\
&+&\frac{1}{2}(\frac{1}{\sqrt{e^{-8\pi\omega(M-D)}+1}})^{p}(\frac{1}{\sqrt{e^{8\pi\omega(M-D)}+1}})^{q}\bigg\{|\overline{0}\rangle\langle\overline{1}|[\bigotimes^{p}_{i=1}(|0_{i}\rangle_{out}\langle1_{i}|)]\notag\\
&\times&[\bigotimes^{q}_{j=1}(|1_{j}\rangle_{in}\langle0_{j}|)]+|\overline{1}\rangle\langle\overline{0}|[\bigotimes^{p}_{i=1}(|1_{i}\rangle_{out}\langle0_{i}|)]\times[\bigotimes^{q}_{j=1}(|0_{j}\rangle_{in}\langle1_{j}|)]\bigg\}\notag\\
&+&\frac{1}{2}|\overline{1}\rangle\langle\overline{1}|\bigotimes^{p}_{i=1}(|1_{i}\rangle_{out}\langle1_{i}|)\bigotimes^{q}_{j=1}(|0_{j}\rangle_{in}\langle0_{j}|).
\end{eqnarray}
Through direct calculation, we get the analytical expression of the coherence of GHZ state for fermionic field as
\begin{eqnarray}\label{S35}
C^{F}(GHZ)=(e^{-8\pi\omega(M-D)}+1)^{-\frac{p}{2}}(e^{8\pi\omega(M-D)}+1)^{-\frac{q}{2}}.
\end{eqnarray}
According to the Eqs.(\ref{S21}), (\ref{S22}) and (\ref{S26}), the wave function of W state for Dirac field can be rewritten as
\begin{eqnarray}\label{S36}
W^{F}_{123\ldots N+n}
&&=\frac{1}{\sqrt{N}}\bigg\{\bigotimes^{n}_{i=1}|1\rangle_{_{N}}|0\rangle_{_{N-1}}\cdots|0\rangle_{n
+1}\big[(e^{-8\pi\omega(M-D)}+1)^{-\frac{1}{2}}|0_{i}\rangle_{out}|0_{i}\rangle_{in}\notag\\
&+&(e^{8\pi\omega(M-D)}+1)^{-\frac{1}{2}}|1_{i}\rangle_{out}|1_{i}\rangle_{in}\big]+\bigotimes^{n}_{i=1}|0\rangle_{_{N}}|1\rangle_{_{N-1}}\cdots|0\rangle_{n+1}\notag\\
&\times&\big[(e^{-8\pi\omega(M-D)}+1)^{-\frac{1}{2}}|0_{i}\rangle_{out}|0_{i}\rangle_{in}+(e^{8\pi\omega(M-D)}+1)^{-\frac{1}{2}}|1_{i}\rangle_{out}|1_{i}\rangle_{in}\big]\notag\\
&+&\cdots+\bigotimes^{n}_{i=2}|0\rangle_{N}|0\rangle_{N-1}\cdots|0\rangle_{n+1}\big[(e^{-8\pi\omega(M-D)}+1)^{-\frac{1}{2}}|0_{i}\rangle_{out}|0_{i}\rangle_{in}\notag\\
&+&(e^{8\pi\omega(M-D)}+1)^{-\frac{1}{2}}|1_{i}\rangle_{out}|1_{i}\rangle_{in}\big]\big[|1_{1}\rangle_{out}|1_{1}\rangle_{in}\big]\bigg\}.
\end{eqnarray}
In a similar way, we consider a general system $\rho^{F,W}_{N-n,p,q}$. Since the density operator of W state for Dirac field is complex, we do not write it out explicitly for simplicity. Through tedious and direct calculations, the coherence of W state between the accessible and inaccessible modes can be calculated as
\begin{eqnarray}\label{S37}
C^{F}(W)&=&\frac{1}{N}\bigg[p(p-1)\big(e^{-8\pi\omega(M-D)}+1\big)^{-1}+q(q-1)\big(e^{8\pi\omega(M-D)}+1\big)^{-1}\notag\\
&+&2p(N-p-q)\big(e^{-8\pi\omega(M-D)}+1\big)^{-\frac{1}{2}}+2q(N-p-q)\big(e^{8\pi\omega(M-D)}+1\big)^{-\frac{1}{2}}\notag\\
&+&(N-p-q)(N-p-q-1)\bigg].
\end{eqnarray}

\begin{figure}
\begin{minipage}[t]{0.5\linewidth}
\centering
\includegraphics[width=3.0in,height=5.2cm]{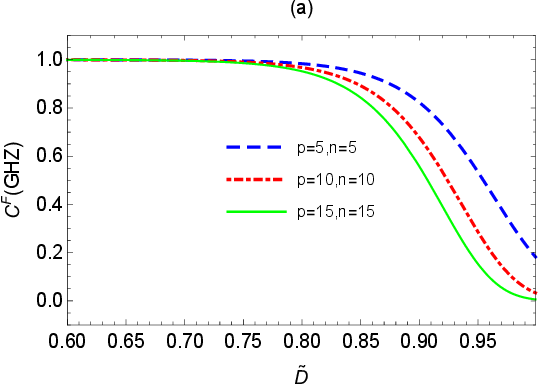}
\label{fig3a}
\end{minipage}%
\begin{minipage}[t]{0.5\linewidth}
\centering
\includegraphics[width=3.0in,height=5.2cm]{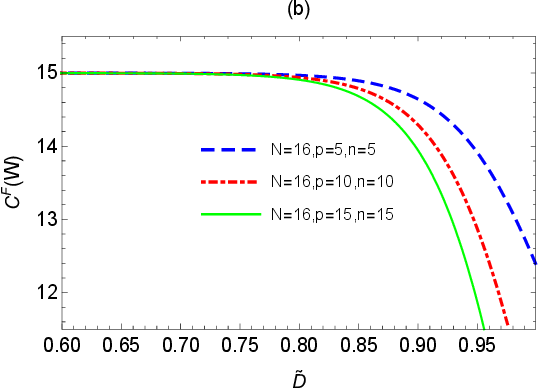}
\label{fig3b}
\end{minipage}%

\begin{minipage}[t]{0.5\linewidth}
\centering
\includegraphics[width=3.0in,height=5.2cm]{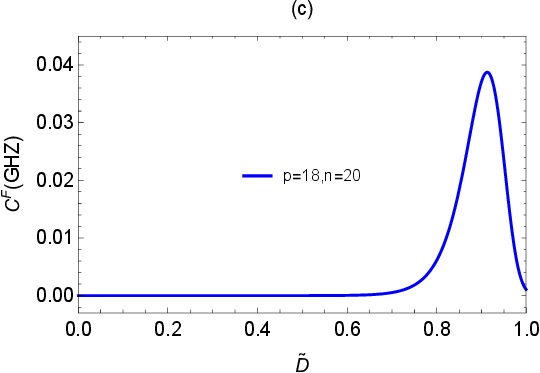}
\label{fig3c}
\end{minipage}%
\begin{minipage}[t]{0.5\linewidth}
\centering
\includegraphics[width=3.0in,height=5.2cm]{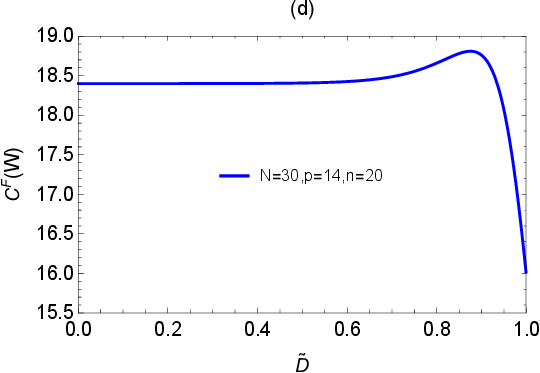}
\label{fig3d}
\end{minipage}%

\begin{minipage}[t]{0.5\linewidth}
\centering
\includegraphics[width=3.0in,height=5.2cm]{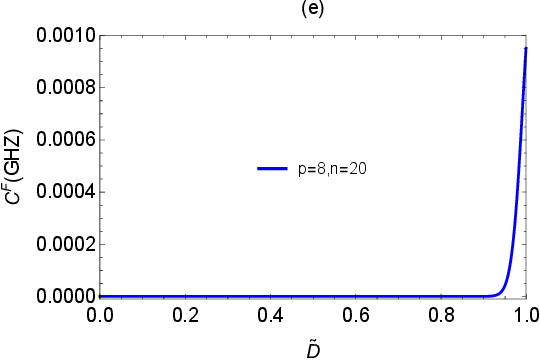}
\label{fig3e}
\end{minipage}%
\begin{minipage}[t]{0.5\linewidth}
\centering
\includegraphics[width=3.0in,height=5.2cm]{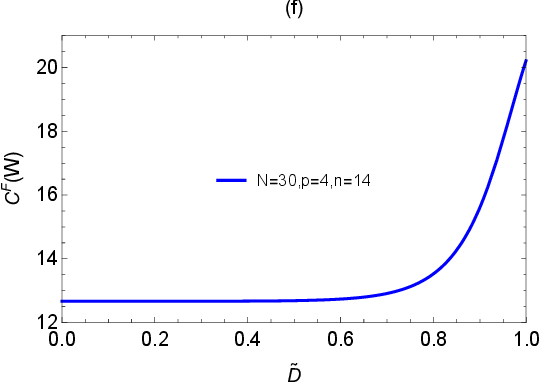}
\label{fig3f}
\end{minipage}%
\caption{Quantum coherences $C^{F}(GHZ)$ and $C^{F}(W)$ for fermionic field as a function of the dilaton $\widetilde{\mathcal{D}}$ for different $N$, $p$, $n$, where we have fixed $M=\omega=1$.}
\label{Fig3}
\end{figure}
Fig.\ref{Fig3} shows how the dilaton $\widetilde{\mathcal{D}}$ of the black hole influences the coherences of GHZ and W states for fermionic field. From Fig.\ref{Fig3}, we can see that the curve of the fermionic field is similar to the curve of the bosonic field in curved spacetime. Through the analysis, we get the conclusions of fermionic coherence: (i) when $p=n$, the N-partite physically accessible coherence first decreases and then reaches to a fixed value with the  dilaton $\widetilde{\mathcal{D}}$; (ii) the inaccessible coherence increases either monotonically or non-monotonically with the dilaton $\widetilde{\mathcal{D}}$, depending on the relative numbers between the accessible and the inaccessible modes; (iii) the fermionic  coherence of  W state is always larger than that of GHZ state.

\begin{figure}
\begin{minipage}[t]{0.5\linewidth}
\centering
\includegraphics[width=3.0in,height=5.2cm]{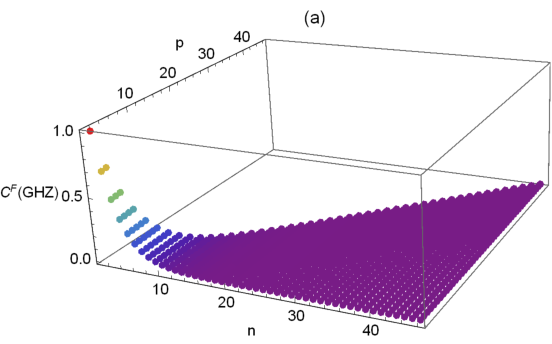}
\label{fig4a}
\end{minipage}%
\begin{minipage}[t]{0.5\linewidth}
\centering
\includegraphics[width=3.0in,height=5.2cm]{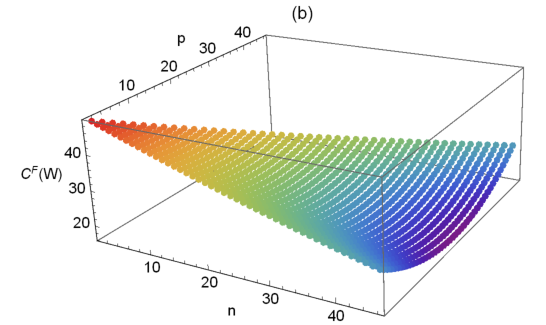}
\label{fig4b}
\end{minipage}%
\caption{Quantum coherences of GHZ and W states for fermionic field as functions of $p$ and $n$ for fixed $M=\omega=1$ and $\widetilde{\mathcal{D}}\longrightarrow1$.}
\label{Fig4}
\end{figure}
Fig.\ref{Fig4} plots the coherences of GHZ and W states for fermionic field as functions of $p$ and $n$ in the limit of  $D\rightarrow M$. Similar with conclusions of the bosonic coherence, we can get three conclusions of fermionic field as follows: (i) for $p=n$ (diagonal line), the physically accessible coherence monotonically decreases with the growth of $p$; (ii) with the increase of $n$, the fermionic coherence of GHZ state monotonically decreases to zero, while the fermionic coherence of W state can always survive forever; (iii) for fixed $n$, with increasing $p$, the fermionic coherence of GHZ state remains unchanged under extremue circumstances $D\rightarrow M$, while the fermionic coherence of W state firstly decreases to minimum and then increases to a fixed value. From Eq.(\ref{S35}), it is note that for $D\rightarrow M$, and we can obtain
\begin{equation}\nonumber
\begin{aligned}
\lim_{ D\rightarrow M}\bigg(e^{-8\pi\omega(M-D)}+1\bigg)^{-\frac{1}{2}}=\lim_{ D\rightarrow M}\bigg(e^{8\pi\omega(M-D)}+1\bigg)^{-\frac{1}{2}}=\frac{\sqrt{2}}{2}.
\end{aligned}
\end{equation}
Therefore, the above analytical expression reflects that the fermionic coherence of GHZ state remains unchanged with the increase of $p$  in the limit of $\widetilde{\mathcal{D}}\rightarrow M$. In this case,  the coherence of GHZ state for fermionic field  can be written in a simpler form
\begin{equation}\nonumber
\begin{aligned}
\lim_{ D\rightarrow M}C^{F}(GHZ)=\lim_{ D\rightarrow M}\bigg(e^{-8\pi\omega(M-D)}+1\bigg)^{-\frac{n}{2}}=\big(\frac{\sqrt{2}}{2}\big)^{n}.
\end{aligned}
\end{equation}

\subsection{Compared coherence with entanglement of bosonic and fermionic fields}
In previous papers, these results have shown that fermionic entanglement is larger than bosonic entanglement in relativistic frame, which means that fermionic entanglement is better suitable for handling relativistic quantum information tasks \cite{L30,L31,L32,L33,L34,L35,L36,L37}. Generally, bosonic entanglement disappears in the case of an extreme black hole, while fermionic entanglement can survive indefinitely \cite{L33,P31,L237,L238,L239,gh1}. Naturally, a question arose: is fermionic coherence larger than bosonic coherence in dilaton spacetime? It is essential to select appropriate quantum resources and types of particles to effectively handle relativistic quantum information tasks in dilaton spacetime. In this section, we will compare coherence with entanglement of bosonic and fermionic fields in curve spacetime.

Due to the complexity of entanglement in the bosonic field, we consider the simplest model: tripartite entanglement. We briefly introduce the measurement of quantum entanglement. The bipartite entanglement between one subsystem and the remaining two subsystems is called one-tangle
\begin{eqnarray}\label{S38}
N_{\alpha(\beta\gamma)}=\|\rho_{\alpha\beta\gamma}^{T_{\alpha}}\|-1.
\end{eqnarray}
Here $T_{\alpha}$ is partial transpose of $\rho_{\alpha\beta\gamma}$ relatives to observer $\alpha$. Note that $\|A\|-1$ is actually equal to the two times of the sum of absolute values of the negative eigenvalues of the operator $A$ \cite{P28,P29}. Thus, one-tangle  can also be expressed as
\begin{eqnarray}\label{S39}
N_{\alpha(\beta\gamma)}=2\sum^{n}_{i=1}|\lambda_{\alpha(\beta\gamma)}^{(-)}|^{i}.
\end{eqnarray}
According to the Eq.(\ref{S39}), we can calculate the one-tangle of GHZ and W states for bosonic and fermionic fields in dilaton spacetime. Here, we consider only one observation outside the event horizon of the black hole, while the rest remain in an asymptotically flat region.  Unfortunately, we cannot write out the analytical expressions of quantum entanglement because they are complex.
\begin{figure}
\begin{minipage}[t]{0.5\linewidth}
\centering
\includegraphics[width=3.0in,height=5.2cm]{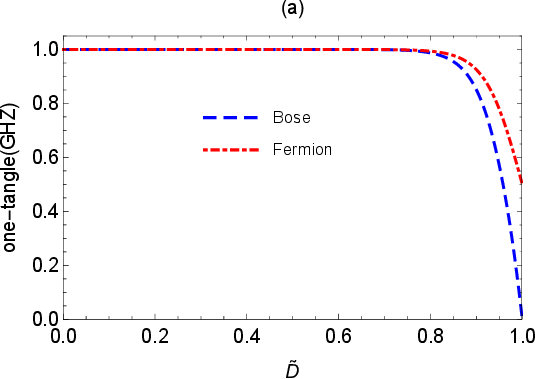}
\label{fig5a}
\end{minipage}%
\begin{minipage}[t]{0.5\linewidth}
\centering
\includegraphics[width=3.0in,height=5.2cm]{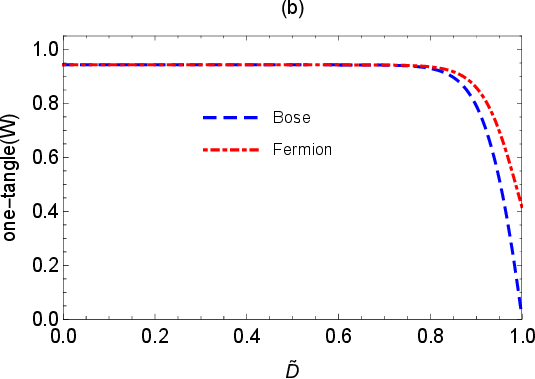}
\label{fig5b}
\end{minipage}%

\begin{minipage}[t]{0.5\linewidth}
\centering
\includegraphics[width=3.0in,height=5.2cm]{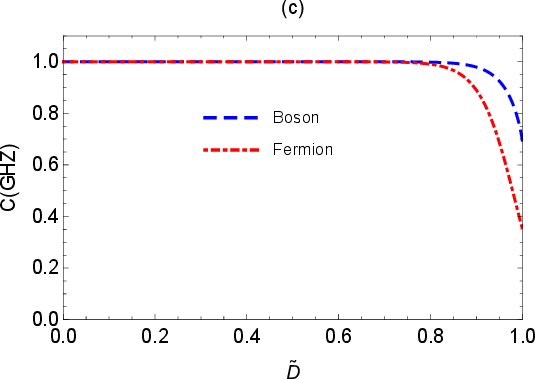}
\label{fig5c}
\end{minipage}%
\begin{minipage}[t]{0.5\linewidth}
\centering
\includegraphics[width=3.0in,height=5.2cm]{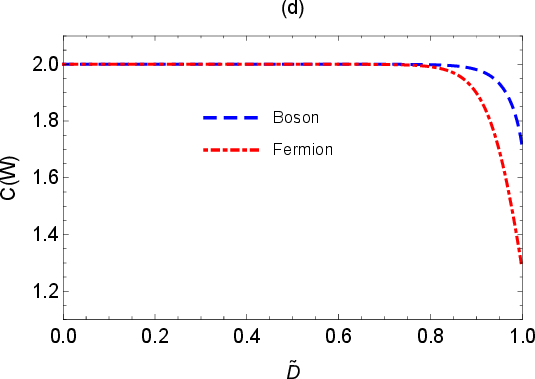}
\label{fig5d}
\end{minipage}%
\caption{Quantum entanglements and coherences of tripartite states for bosonic and fermionic fields as a function of the dilaton $\widetilde{\mathcal{D}}$ for fixed $M=\omega=1$.}
\label{Fig5}
\end{figure}

Fig.\ref{Fig5} shows how the dilaton $\widetilde{\mathcal{D}}$ of the black hole affects quantum entanglements and coherences of GHZ and W states for bosonic and fermionic fields. From Fig.\ref{Fig5}(a) and (b), we can see that for GHZ or W state, fermionic entanglement is always larger than bosonic entanglement. Comparing Fig.\ref{Fig5}(a) with (b), we find that quantum entanglement of GHZ state is larger than
quantum entanglement of W state in dialton spacetime. Unlike quantum entanglement in dilaton spacetime, Fig.\ref{Fig5}(c) and (d) show that for GHZ or W state, bosonic coherence  is larger than fermionic coherence. However, Fig.\ref{Fig5}(c) and (d) also show that the coherence of W state is larger than the coherence of GHZ state. Through the above analysis, the properties of quantum entanglements and  coherences of GHZ and W states for bosonic and fermionic fields are completely opposite. Hence, some conclusions in the dilaton black hole can be obtained as follow: (i) fermionic entanglement is more suitable for processing relativistic quantum information compared to bosonic entanglement; (ii) bosonic coherence is more effective for processing relativistic quantum information than fermionic coherence; (iii) quantum entanglement of GHZ state is a superior quantum resource compared to quantum coherence of W state (iv)
the coherence of W state is a better quantum resource than the coherence of GHZ state.
In other words, for different types of particles, we should choose appropriate quantum resources to deal with quantum information tasks.
\section{ Conclutions  \label{GSCDGE}}

In this paper, we have studied N-partite coherences of GHZ and W states for bosonic and fermionic fields in a GHS dilaton black hole. We initially assume that $N$ observes share GHZ and W states at the same point in the asymptotically flat region of the dilaton black hole. Then, $N-n$ observers remain stationary in the asymptotically flat region, while $n$ $(1\leq n\leq N)$ observers hover near the event horizon of the black hole. We obtain the general analytical expressions for N-partite coherence that include all physically accessible and inaccessible N-partite coherences in the dilaton black hole. We can find that the coherence of GHZ state is independs of the number of initial particles $N$, while the coherence of W state is dependent on the number of initial particles $N$. The inaccessible N-partite coherence increases either monotonically or non-monotonically with the increase of the dilaton, depending on the relative numbers between the accessible and the inaccessible modes.  Interestingly, the coherences of GHZ state for bosonic and fermionic fields is independs of the relative numbers between the accessible and the inaccessible modes in the case of an extreme black hole. Additionally, we obtain four key conclusions: (i) quantum coherence of W state is always larger than quantum coherence of GHZ state; (ii) quantum entanglement of GHZ state is always larger than quantum entanglement of W state;
(iii) for GHZ or W state,  bosonic coherence is larger than  fermionic coherence; (iv) fermionic entanglements is larger than bosonic entanglement for GHZ or W state.  Therefore, we should select the appropriate quantum resources to handle relativistic quantum information tasks. We hope that this research will help us understand how quantum information travels both inside and outside the event horizon of the black hole.

\begin{acknowledgments}
This work is supported by the National Natural
Science Foundation of China (Grant Nos. 12205133), LJKQZ20222315,  and  the Special Fund for Basic Scientific Research of Provincial Universities in Liaoning under grant NO. LS2024Q002. The authors would like
to thank Wentao Liu for helpful discussions.
\end{acknowledgments}
$\textbf{Data Availability Statement}$

This manuscript has no associated data or
the data will not be deposited. [Authors comment: All data included in
this study are available upon request by contact with the corresponding
author Shu-Min Wu.]

\appendix
\onecolumngrid
%


\end{document}